%% file: main.tex
\documentclass[11pt]{article}

\usepackage[utf8]{inputenc}
\usepackage[a4paper,margin=1in]{geometry}
\usepackage{microtype}
\usepackage{amsmath,amssymb}
\usepackage{booktabs,tabularx,array}
\usepackage[table]{xcolor}
\usepackage{graphicx}
\usepackage{caption}
\usepackage{enumitem}
\usepackage[numbers,sort&compress]{natbib}
\usepackage{tikz}
\usetikzlibrary{arrows.meta,positioning,calc,fit,backgrounds,shapes.geometric}
\usepackage{pgfplots}
\pgfplotsset{compat=1.18}
\usepgfplotslibrary{groupplots,fillbetween}
\usepackage[hidelinks]{hyperref}
\usepackage[capitalise,noabbrev]{cleveref}

\definecolor{accent}{HTML}{2A78D6}
\definecolor{accentlight}{HTML}{D4E4F7}
\definecolor{neutral}{HTML}{B9B8B2}
\definecolor{ink}{HTML}{0B0B0B}
\definecolor{muted}{HTML}{52514E}
\definecolor{rule}{HTML}{E4E3DF}

\setlist{itemsep=2pt,topsep=4pt}
\newcolumntype{L}{>{\raggedright\arraybackslash}X}
\newcolumntype{P}[1]{>{\raggedright\arraybackslash}p{#1}}

\hypersetup{
  pdftitle={Mitigating Risks in Software Development through Effective Requirements Engineering},
  pdfauthor={Valentin Burkin},
  pdfkeywords={requirements engineering, requirements elicitation, software lifecycle, software risk, security requirements}
}

\title{Mitigating Risks in Software Development\\through Effective Requirements Engineering%
\thanks{This is a revised version (v2) of the preprint first posted in May 2023. The language,
structure, figures and citations have been revised and several factual statements corrected;
the scope and the main arguments are unchanged.}}
\author{Valentin Burkin\\[2pt]
\small The Grainger College of Engineering, University of Illinois Urbana-Champaign, USA\\
\small\texttt{vburkin2@illinois.edu}}
\date{}

\begin{document}
\maketitle

\begin{abstract}
\noindent
Problems that originate in requirements are common and expensive in software projects,
and they are a recurring cause of late delivery, budget overruns and cancellation. This paper is a narrative overview of requirements engineering
(RE) as a means of reducing these risks. It outlines the core RE activities (elicitation,
analysis and negotiation, specification, validation and change management) and the kinds
of statements a requirements document contains, including the distinction between
requirements, domain properties, assumptions and definitions. It reviews the evidence on the
cost of requirements defects, together with the main criticisms of that evidence, and
compares how the waterfall, spiral, iterative (Rational Unified Process) and agile lifecycles
treat requirements. It then discusses stakeholder identification, elicitation techniques and
common obstacles to elicitation, and describes practical ways of turning vague or
conflicting stakeholder statements, including security and performance expectations, into
precise, verifiable and prioritized requirements. The paper is intended as an introduction
for practitioners and students who work with requirements but are not RE specialists.
\end{abstract}

\medskip
\noindent\textbf{Keywords:} requirements engineering; requirements elicitation; software
requirements specification; software lifecycle models; security requirements; software
project risk

% ===========================================================================
\section{Introduction}
\label{sec:intro}

A software system can be built correctly and still fail if it solves the wrong problem.
Requirements engineering (RE) is the part of software engineering concerned with
discovering what a system should do, the qualities it should have, and the constraints
under which it must operate, and with documenting and maintaining that understanding as
the project evolves \citep{nuseibeh2000,vanlamsweerde2009}. Because almost every later
activity, including architecture, implementation, testing, security review and
documentation, depends on the requirements, errors made at this stage propagate through
the whole project.

The people who carry out RE work are often called requirements engineers or business
analysts. They act as the link between the customer community and the development team
\citep{wiegers2013}. Their output is used by many other roles: developers and architects
design against it, testers derive test cases from it, product and project managers plan
releases with it, technical writers document the system from it, and security engineers
use it to identify what must be protected. In small teams, developers often write some or all of
the requirements themselves. A working knowledge of RE is therefore useful well
beyond the specialist role.

Security is a good illustration of why requirements matter. Security properties such as
confidentiality, integrity, availability and accountability depend on knowing which assets
must be protected, against which threats, and under which assumptions about the
environment \citep{haley2008}. If these are not identified and agreed on while requirements
are being defined, important threats may be left unaddressed, or controls may be added that
conflict with usability and performance expectations \citep{firesmith2003}.

This paper is a narrative overview rather than a systematic literature review. Its aim is to
give readers who are not RE specialists a compact account of the main activities, concepts
and trade-offs, with pointers to the primary literature. The paper is organized as follows.
\Cref{sec:process} describes the RE process and the kinds of statements a requirements
document contains. \Cref{sec:cost} reviews the evidence on the cost of requirements defects.
\Cref{sec:lifecycles} compares how common lifecycle models treat requirements.
\Cref{sec:elicitation} covers stakeholder identification and elicitation.
\Cref{sec:analysis} addresses vague and conflicting requirements, security trade-offs and
prioritization. \Cref{sec:spec} covers specification, validation and change management, and
\cref{sec:conclusion} concludes.

% ===========================================================================
\section{Requirements Engineering: Process and Artifacts}
\label{sec:process}

\subsection{Core activities}

RE is usually described as a set of interleaved activities rather than a single phase
\citep{nuseibeh2000,vanlamsweerde2009,wiegers2013}. \Cref{fig:process} summarizes them.

\begin{description}[leftmargin=0pt,labelindent=0pt,font=\normalfont\itshape]
  \item[Domain understanding and elicitation.] The analyst studies the organization, the
  application domain and any existing system, and gathers needs, constraints and
  background knowledge from stakeholders and from existing documents.
  \item[Analysis and negotiation.] The elicited statements are checked for conflicts,
  gaps and ambiguity. Alternatives are evaluated, conflicts are resolved with the
  stakeholders concerned, and requirements are prioritized.
  \item[Specification.] The agreed requirements are documented, typically in a software
  requirements specification (SRS), a set of use cases, or a product backlog of user
  stories with acceptance criteria.
  \item[Validation.] Stakeholders and the development team check that the documented
  requirements describe what is actually needed, and that each one can be implemented
  and verified.
  \item[Requirements management.] Throughout the project, changes to the requirements are
  controlled, versions are kept, and each requirement is traced to its source and to the
  design elements and tests that realize it.
\end{description}

\begin{figure}[t]
  \centering
  \input{figures/process.tex}
  \caption{Requirements engineering activities. The four core activities are performed
  iteratively; validation findings and change requests feed back into elicitation.
  Requirements management spans the whole project. Adapted from the process descriptions
  in \citep{nuseibeh2000,vanlamsweerde2009}.}
  \label{fig:process}
\end{figure}

The activities are iterative. Analysis frequently reveals questions that require further
elicitation, and validation often exposes misunderstandings that send the analyst back to
the stakeholders. The amount of documentation produced also varies widely. Projects run
under heavily documented processes may produce specifications of hundreds of pages, whereas
agile teams may maintain only a backlog and acceptance tests
(\cref{sec:lifecycles}). In both cases the underlying activities are the same.

\subsection{Kinds of statements in a requirements document}
\label{sec:statements}

A requirements document contains more than requirements. Following \citet{zave1997} and
\citet{vanlamsweerde2009}, it is useful to distinguish \emph{prescriptive} statements, which
state what should be made true, from \emph{descriptive} statements, which state what is
true of the environment regardless of the system, and to note which part of the world is
responsible for each. \Cref{tab:statements} lists the main types.

\begin{table}[t]
  \caption{Types of statement in a requirements document, after
  \citet{zave1997} and \citet{vanlamsweerde2009}. The train examples follow
  \citet{vanlamsweerde2009}; the meeting-scheduler examples are illustrative.}
  \label{tab:statements}
  \small
  \begin{tabularx}{\textwidth}{@{}P{2.55cm}P{2.0cm}LL@{}}
    \toprule
    Statement type & Mood & Purpose & Example \\
    \midrule
    System requirement & Prescriptive &
      Desired property of the system as a whole, stated in terms of phenomena in the
      environment &
      The train doors shall remain closed while the train is moving. \\
    Software requirement & Prescriptive &
      Property the software alone must enforce, stated in terms of its inputs and
      outputs &
      The output \texttt{doorsState} shall be \texttt{closed} whenever the input
      \texttt{measuredSpeed} is non-zero. \\
    Domain property & Descriptive &
      Fact about the environment that holds whatever the system does, often a physical
      law or organizational fact &
      A train is moving if and only if its physical speed is non-zero. A person cannot
      attend two meetings at the same time. \\
    Assumption & Prescriptive, for the environment\textsuperscript{a} &
      Property that some component of the environment (a device, a person, another
      system) is expected to satisfy; the software cannot enforce it &
      The speed sensor reports a non-zero value exactly when the train is moving.
      Participants answer meeting requests within three days. \\
    Definition & --- &
      Meaning of a term used in the document; has no truth value &
      A person \emph{participates} in a meeting if they attend it from beginning to end. \\
    \bottomrule
  \end{tabularx}

  \smallskip
  \footnotesize\textsuperscript{a}\,\citet{vanlamsweerde2009} treats assumptions as
  generally prescriptive statements on environment components; \citet{zave1997} include them
  in descriptive domain knowledge.
\end{table}

The distinction matters because a software requirement satisfies a system requirement only
in combination with the domain properties and assumptions. In the train example
(\cref{fig:worldmachine}), the software requirement guarantees that the door-control
output is \texttt{closed} whenever the measured speed is non-zero. This ensures that the
doors stay closed while the train moves only if the speed sensor and the door actuators
behave as assumed. Writing this down,
\begin{equation}
  \textsc{SofReq},\ \textsc{Asm},\ \textsc{Dom} \;\models\; \textsc{SysReq},
  \label{eq:satisfaction}
\end{equation}
makes the assumptions explicit, so that they can be checked with the stakeholders who own
the sensor and the actuators \citep{zave1997,vanlamsweerde2009}. Failures, including security
failures, can arise when an assumption was never stated and does not hold in practice
\citep{vanlamsweerde2009,haley2008}.

Assumptions also influence scope. In a meeting scheduler, if it is assumed that
participants reply promptly to requests for their availability, the system can simply wait
for replies. If that assumption cannot be justified, the system needs an additional
requirement, such as automatic reminders. Definitions deserve similar care. Whether
\emph{participation} means attending the whole meeting, part of it, or simply joining
remotely changes what the scheduler must record.

\begin{figure}[t]
  \centering
  \input{figures/worldmachine.tex}
  \caption{System requirement, software requirement, domain property and assumptions in
  the train door example of \citet{vanlamsweerde2009}. The software can observe and control
  only the shared phenomena at the interface. The system requirement, stated in terms of
  the environment, holds when the software requirement, the domain property and the
  assumptions about the sensor and the actuator hold together (\cref{eq:satisfaction}).}
  \label{fig:worldmachine}
\end{figure}

\subsection{Functional and non-functional requirements}

Orthogonal to the classification above, requirements are commonly divided into
\emph{functional} requirements, which describe the behavior the system must provide, and
\emph{non-functional} requirements, which describe qualities such as performance,
reliability, usability, security and maintainability, or constraints on the solution
\citep{glinz2007}. Non-functional requirements are harder to elicit and to verify because
stakeholders tend to express them in general terms (``fast'', ``secure'', ``easy to use''),
and for most qualities other than performance there are no generally agreed measures
\citep{glinz2007}.
\Cref{sec:vague} returns to this problem.

It is also helpful to separate \emph{business requirements}, which state the objectives the
organization wants to achieve with the system, from \emph{business rules}, which are
policies, guidelines, standards, regulations and computations that constrain the business
and exist independently of any particular application \citep{wiegers2013}. Business rules frequently originate
outside the project, for example in legislation, and are often the source of security and
privacy requirements.

\subsection{The software requirements specification}

Whatever form it takes, the SRS serves several purposes. It is the basis of agreement
between customers and developers on what will be built, the reference against which
the system is designed, tested and accepted, and the starting point for later maintenance
and enhancement. It typically states the purpose and scope of the system, its functional
requirements, its non-functional requirements, its external interfaces (user interfaces,
APIs, hardware and other systems), and the constraints, assumptions and definitions that
apply \citep{ieee830,iso29148}.

IEEE Std 830 lists eight characteristics of a good SRS \citep{ieee830}; its successor
ISO/IEC/IEEE 29148 covers similar ground, separating characteristics of individual
requirements from characteristics of a requirement set \citep{iso29148}. \Cref{tab:quality}
restates them as questions a reviewer can ask.

\begin{table}[t]
  \caption{Quality characteristics of a software requirements specification
  \citep{ieee830}, phrased as review questions.}
  \label{tab:quality}
  \small
  \begin{tabularx}{\textwidth}{@{}P{3.0cm}L@{}}
    \toprule
    Characteristic & Review question \\
    \midrule
    Correct & Does every requirement state something the system is actually required to do? \\
    Unambiguous & Does every requirement have exactly one reasonable interpretation? \\
    Complete & Are all significant requirements, responses to all inputs, and all terms covered? \\
    Consistent & Is it free of requirements that contradict one another? \\
    Ranked & Is each requirement marked with its importance and expected stability? \\
    Verifiable & Can each requirement be checked by a finite, cost-effective test, inspection or analysis? \\
    Modifiable & Can a requirement be changed in one place without unintended side effects? \\
    Traceable & Is the origin of each requirement clear, and can it be referenced by design, code and tests? \\
    \bottomrule
  \end{tabularx}
\end{table}

% ===========================================================================
\section{The Cost of Requirements Defects}
\label{sec:cost}

\subsection{Evidence from project surveys}

The most frequently cited evidence on project failure is the CHAOS report published by the
Standish Group in 1994 \citep{standish1994}. Based on a survey of 365 IT executives
covering 8{,}380 applications, it classified 16.2\% of projects as successful (on time, on
budget and with all specified features), 52.7\% as \emph{challenged} (completed and operational,
but over budget, late and with fewer features than specified) and 31.1\% as \emph{impaired} (canceled during
development). Challenged and impaired projects together reported an average cost of 189\%
of the original estimate, and challenged projects delivered on average 61\% of the
originally specified features.

Respondents were also asked about the causes of trouble. As \cref{fig:chaos} shows, lack of
user input and incomplete or changing requirements were the three most frequently named
factors for challenged projects, and incomplete requirements and lack of user involvement
were the two most frequently named factors for impaired projects. In other words, the
factors most often blamed for project problems are requirements problems.

\begin{figure}[t]
  \centering
  \input{figures/chaos.tex}
  \caption{Results of the 1994 CHAOS survey \citep{standish1994}. (a) Distribution of
  project outcomes. (b, c) The five factors most often named by respondents for challenged
  and impaired projects, as a percentage of responses; factors related to requirements or
  user involvement are highlighted. See \cref{sec:cost} for criticism of these figures.}
  \label{fig:chaos}
\end{figure}

These figures should be read with caution. \Citet{jorgensen2006} found that the 189\% cost
overrun is much higher than the overruns reported by other studies of the period, and that
the way the survey selected and questioned respondents was likely to over-represent
problem projects. \Citet{eveleens2010} argued that the Standish definition of success
measures only how closely a project met its initial estimates of cost, time and
functionality, so that it reflects the quality of estimation as much as the outcome of the
project. The
exact percentages are therefore unreliable. The qualitative finding that requirements and
user involvement are among the leading causes of project trouble is, however, consistent
with later field studies. \Citet{hall2002} analyzed the requirements problems reported in twelve
software companies and found most of them to be organizational rather than technical. In a
survey of 228 companies, \citet{mendez2017} found incomplete or hidden requirements,
communication flaws between the team and the customer, and moving targets to be the most
frequently reported RE problems, and respondents often linked them to project failure.

\subsection{Cost escalation}

A second argument for investing in RE concerns the cost of correcting a defect at
different points of the lifecycle. \Citet{boehm1981} reported that the cost of fixing a
requirements defect rises steeply the later it is found, and \citet{boehm2001} summarized
the evidence by stating that a problem found and fixed after delivery is often about 100
times more expensive than one found during requirements and design, and closer to 5 times
for small, non-critical systems. More recent data are less clear-cut. \Citet{menzies2017} analyzed 171 projects that
followed the Team Software Process and found, in most cases, no significant increase in the
time needed to resolve issues that were fixed later. The size of the escalation therefore
depends on the process and the kind of system. It seems reasonable to expect it to be
largest when a requirements error has become embedded in the architecture, in contracts or
in deployed data, although this is a judgement rather than a measured result.

The practical conclusion is unchanged. Requirements errors are generally cheaper to remove before
design and implementation have been built on them, so RE processes should include explicit
reviews, prototypes and other checks that expose such errors early
(\cref{sec:validation}).

% ===========================================================================
\section{Requirements Across Lifecycle Models}
\label{sec:lifecycles}

The lifecycle model chosen for a project determines when requirements are elicited, how
much of them is written down, and how changes are absorbed. \Cref{fig:lifecycles} sketches
the distribution of RE effort over time in four common models, and \cref{tab:lifecycles}
compares them.

\begin{figure}[t]
  \centering
  \input{figures/lifecycles.tex}
  \caption{Schematic distribution of requirements engineering effort over the duration of a
  project in four lifecycle models. The curves are illustrative, not measured data. The RUP
  panel approximates the requirements discipline in the RUP overview chart
  \citep{kruchten2004}, using a typical 10/30/50/10\% split of the phases Inception (I),
  Elaboration (E), Construction (C) and Transition (T).}
  \label{fig:lifecycles}
\end{figure}

\subsection{Waterfall}

The waterfall model arranges requirements, design, implementation, testing and maintenance
as consecutive phases. It is usually attributed to \citet{royce1970}, although Royce himself
described the purely sequential version as risky and inviting failure, and recommended
iteration with neighboring steps and building an early pilot version of the system (``do it
twice''). In its sequential form, requirements are elicited and documented in
full at the start, then frozen and placed under formal change control. The approach gives
clear milestones and comprehensive documentation, and it suits projects with stable,
well-understood requirements and contractual or regulatory demands for documentation. Its
weakness is that testing comes late, so requirements errors are discovered when they are
most expensive to fix, and changes in the customer's needs, technology or circumstances
during a long project are difficult to absorb.

\subsection{Spiral}

The spiral model \citep{boehm1988} organizes development as a series of cycles, each of
which determines objectives and alternatives, identifies and resolves the main risks
(often by building a prototype), develops and verifies the next increment, and plans the
next cycle. Requirements are revisited in every cycle, and the most uncertain requirements
can be explored with prototypes before the team commits to them. The model is particularly
useful for large, complex projects with high technical or managerial risk, and where early
user feedback through prototypes is valuable. Its costs are the
effort and expertise needed for continuous risk analysis, and greater difficulty in
contracting and planning than with a sequential model.

\subsection{Iterative: the Rational Unified Process}

The Rational Unified Process (RUP), developed at Rational Software in the late 1990s
\citep{kruchten2004}, divides a project into four phases (Inception, Elaboration,
Construction and Transition), each consisting of one or more iterations. Activities are
grouped into disciplines, such as requirements, analysis and design, implementation and
test, and every iteration contains some work from most disciplines. The balance changes
over the project. In Elaboration most effort goes into refining the requirements and
establishing a stable architecture; in Construction the emphasis moves to implementation and
testing, while requirements work continues at a lower level (\cref{fig:lifecycles}).

RUP accepts that requirements change. Instead of freezing them, it aims to understand the
most important requirements well enough to build a robust architecture early and then to
refine the rest iteration by iteration. Careful prioritization means that each iteration
delivers the features of highest value, so even an incomplete release meets part of the
customer's objectives.

\subsection{Agile development}

The Manifesto for Agile Software Development \citep{beck2001}, written in 2001 by seventeen
practitioners, values individuals and interactions over processes and tools, working
software over comprehensive documentation, customer collaboration over contract negotiation,
and responding to change over following a plan. For RE, this means continuous
face-to-face communication with a customer representative, requirements expressed as user
stories with acceptance criteria, short iterations that turn requirements into working
software quickly, and a backlog that is reprioritized continually
\citep{paetsch2003,ramesh2010,delucia2010,hamed2013}.

Agile RE reduces the risk of building the wrong product, because frequent delivery and
customer feedback expose misunderstandings early \citep{paetsch2003,ramesh2010}. It has its own risks, however. \Citet{ramesh2010} report that agile
teams struggle with non-functional requirements, which are hard to express as user stories,
with inadequate architecture when quality requirements are discovered late, and with
dependence on customer access and participation, including the availability of customer
representatives and agreement among them. They also report problems caused by minimal
documentation, which matters in particular when regulators, auditors or future maintainers
need a
written record of what the system is required to do.

\begin{table}[t]
  \caption{Treatment of requirements in four lifecycle models, summarized from the sources
  cited for each model and \cref{sec:lifecycles}.}
  \label{tab:lifecycles}
  \small
  \begin{tabularx}{\textwidth}{@{}P{1.65cm}LLL@{}}
    \toprule
    Model & Requirements treatment & Strengths & Weaknesses \\
    \midrule
    Waterfall \citep{royce1970} &
      Elicited and documented in full up front, then baselined; changes go through formal
      control. &
      Clear milestones; comprehensive documentation; suits stable requirements and regulated
      contexts. &
      Errors found late in testing; poor fit for changing needs; long time to first
      feedback. \\
    Spiral \citep{boehm1988} &
      Revisited in each risk-driven cycle; uncertain requirements explored with
      prototypes. &
      Explicit risk management; early user feedback; incremental commitment. &
      Requires risk-analysis expertise; harder to plan and contract; process
      overhead. \\
    RUP \citep{kruchten2004} &
      Most important requirements refined early to stabilize the architecture; others
      refined per iteration. &
      Balances documentation and flexibility; early architecture; value-driven
      iterations. &
      Process is heavy if not tailored; depends on disciplined iteration
      planning. \\
    Agile \citep{beck2001,ramesh2010} &
      Continuous; user stories in a reprioritized backlog; acceptance tests as
      specification. &
      Fast feedback; responsive to change; close customer collaboration. &
      Weak on non-functional requirements; relies on customer availability; thin written
      record. \\
    \bottomrule
  \end{tabularx}
\end{table}

% ===========================================================================
\section{Eliciting Requirements}
\label{sec:elicitation}

\subsection{Understanding the domain and the current system}

Elicitation begins with domain understanding \citep{vanlamsweerde2009}. Before asking
stakeholders what the new system should do, the analyst needs to understand the
organization (its structure, business objectives, policies and the responsibilities of its
members), the application domain (its concepts, objectives and regulations) and, if there
is one, the system currently in use, whether automated or manual: its objectives, actors,
resources, tasks and workflows, and the problems that motivate the change. Even
stakeholders with years of experience rarely know every part of their domain, so this step
also prepares the analyst to ask informed questions.

Two products of this phase are worth keeping. The first is a preliminary description of the
system as it is, how it fits into the organization and domain, the problems identified, the
opportunities to be exploited and possible alternative solutions. The second is a glossary
of domain terms, started at the beginning and appended to the requirements document, which
reduces misunderstanding between stakeholders who use different vocabulary.

\subsection{Identifying stakeholders}

Stakeholders are the individuals or organizations that stand to gain or lose from the
success or failure of the system \citep{nuseibeh2000}. Typical stakeholders include end users, the
customer who pays for the system, managers who set the project's direction, operators,
maintainers, regulators and people who are affected by the system without using it. It is
usually impossible to involve every stakeholder, so the analyst selects a representative
sample according to their role, their stake in the system and the kind of knowledge they
can provide \citep{vanlamsweerde2009}.

Two points deserve attention. First, the people who make decisions about the requirements
are not always the people who fund the project, and both must be identified. Second,
stakeholders' interests conflict. In a typical business system, management may emphasize
cost, the security officer may emphasize protection of data, and end users may emphasize
speed and ease of use. Identifying stakeholders at every relevant level and area of the
organization reduces the risk that the requirements reflect only the views of those who are
most accessible or most vocal.

\subsection{Elicitation techniques}

Many elicitation techniques exist, and they are complementary
\citep{goguen1993,zowghi2005}. \Cref{tab:techniques} summarizes the most common ones. In a
systematic review of empirical studies, \citet{davis2006} found structured interviews to be
among the most effective techniques for gathering information. In practice, analysts
combine techniques:
interviews and observation for depth, workshops for reconciling views, surveys for breadth,
and prototypes to validate their understanding.

\begin{table}[t]
  \caption{Common elicitation techniques, compiled from
  \citet{goguen1993}, \citet{zowghi2005} and \citet{wiegers2013}.}
  \label{tab:techniques}
  \small
  \begin{tabularx}{\textwidth}{@{}P{2.9cm}LL@{}}
    \toprule
    Technique & Most useful for & Limitations \\
    \midrule
    Interviews &
      Detailed, qualitative understanding of individual stakeholders' goals, tasks and
      problems. &
      Time-consuming; results depend on interviewer skill; tacit knowledge often not
      verbalized. \\
    Workshops and focus groups &
      Collecting and reconciling the views of several stakeholders; building shared
      understanding. &
      Dominant participants can bias outcomes; scheduling is difficult. \\
    Questionnaires and surveys &
      Quantitative data from many, possibly distributed, stakeholders; confirming patterns. &
      Limited depth; answers bounded by the questions asked; low response rates. \\
    Observation and ethnography &
      Understanding actual work practice, including tacit knowledge stakeholders cannot
      articulate. &
      Slow; observer presence may change behavior; hard to generalize. \\
    Document and artifact analysis &
      Business rules, regulations, data formats and features of existing systems. &
      Documents may be outdated or inconsistent with practice. \\
    Prototyping &
      Validating understanding and exploring uncertain requirements, especially user
      interfaces. &
      Stakeholders may treat a prototype as the product; risk of premature design
      decisions. \\
    Consulting subject-matter experts &
      Technical accuracy and feasibility in specialized domains. &
      Experts may not represent end users' needs. \\
    \bottomrule
  \end{tabularx}
\end{table}

\subsection{Obstacles to elicitation}
\label{sec:obstacles}

Elicitation is a social process, and most of its difficulties are communication problems
rather than technical ones \citep{alrawas1996,goguen1993}. Common obstacles include:
\begin{itemize}
  \item \emph{Access.} Key stakeholders are busy, geographically distant, in other time
  zones, or not convinced that their input is worth the effort.
  \item \emph{Tacit knowledge.} Experts omit information they consider common sense, and
  some knowledge of how work is done is not available to conscious reflection at all.
  \item \emph{Articulation.} Stakeholders may find it hard to express their needs, may
  describe solutions rather than problems, or may hold unrealistic expectations of what
  is feasible.
  \item \emph{Vocabulary and background.} Differences in professional language, education
  and culture lead the analyst and stakeholders to understand the same words differently.
  \item \emph{Conflict and politics.} Departments compete, stakeholders have personal
  objectives or resist the change the system will bring, and decisions are influenced by
  organizational power.
  \item \emph{Volatility.} Organisational change, time and cost pressure, and outdated
  documents cause the requirements to shift during elicitation.
  \item \emph{Assumptions.} When information is missing, both analysts and developers
  tend to fill the gap with guesses, which later surface as defects or as work outside the
  agreed scope.
\end{itemize}

The analyst's main tools against these obstacles are interpersonal: listening carefully
and identifying the key points, using the stakeholders' vocabulary rather than technical
jargon, being transparent about how the information will be used, and maintaining a
relationship of trust. Information should be confirmed from more than one source where
possible, and the analyst's integrated understanding of the problem should be presented
back to stakeholders in regular reviews, using diagrams, examples or prototypes to make it
concrete. These reviews also show stakeholders that their contributions are used, which
helps keep them engaged.

\subsection{The role of the analyst}

The analyst gathers, analyses, documents and validates requirements, and manages them as
they change, new insights emerge or new risks appear \citep{wiegers2013}. This involves
deciding whom to talk to, when, why and with which technique; exploring alternative
solutions and discussing their costs and benefits with stakeholders; and writing the
specification in whatever forms suit the audience, including use cases, natural-language
requirements, analysis models, tables, prototypes and, where needed, formal notations.
Although the analyst does not need to be a developer, understanding design and
implementation helps: developers can point out the constraints and capabilities of the
technology, and requirements that ignore these are likely to be infeasible.

% ===========================================================================
\section{Analysis, Negotiation and Prioritization}
\label{sec:analysis}

\subsection{From vague statements to verifiable requirements}
\label{sec:vague}

Stakeholders frequently state requirements in terms that cannot be verified, such as
``the system must be fast'', ``it must never go down'' or ``it must be highly secure''.
Such statements are useful starting points, but they cannot be designed or tested against,
and they hide conflicts: a stakeholder who asks for high security may not accept additional
authentication steps, and a system that must never go down may be unaffordable. The
analyst's task is to ask what the stakeholder actually means, and to replace each vague
statement with one or more measurable requirements that the stakeholder agrees with.
\Cref{tab:vague} gives examples; the figures in it are illustrative and would be set with
the stakeholders.

\begin{table}[t]
  \caption{Examples of vague stakeholder statements and verifiable replacements. Values are
  illustrative.}
  \label{tab:vague}
  \small
  \begin{tabularx}{\textwidth}{@{}P{3.3cm}L@{}}
    \toprule
    Stakeholder statement & Verifiable requirement \\
    \midrule
    ``The system must be fast.'' &
      For up to 500 concurrent users, 95\% of search requests shall return results within
      1.5\,s, measured at the server. \\
    ``It must always be available.'' &
      The booking service shall be available 99.9\% of the time in each calendar month,
      excluding announced maintenance windows of at most 4\,h. \\
    ``It must be very secure.'' &
      Access to patient records shall require multi-factor authentication; every read of a
      record shall be logged with user, time and record identifier, and logs shall be
      retained for 6 years. \\
    ``It should be easy to use.'' &
      After a 15-minute introduction, 90\% of first-time clerks shall complete a standard
      booking without assistance in under 3\,min. \\
    \bottomrule
  \end{tabularx}
\end{table}

\subsection{Conflicts and the security trade-off}

Conflicts between requirements are normal. They arise between stakeholders with different
interests, between quality attributes, and between requirements and business rules. The
analyst makes the conflict explicit, identifies alternative solutions, and negotiates a
resolution with the stakeholders concerned. Possible resolutions include choosing one
requirement over another, weakening one or both, or restricting each to the circumstances
in which it matters most \citep{vanlamsweerde2009}.

Security requirements are a common source of such conflicts, because security controls
often cost performance, usability or money. \Citet{firesmith2003} recommends stating
security requirements as the protection the system must provide rather than as particular
security mechanisms, so that architectural decisions are not made prematurely and the
design remains free to choose an adequate control. Several methods make the process
systematic. SQUARE \citep{mead2005} defines nine steps, from agreeing on definitions and
security goals through risk assessment, elicitation, categorization and prioritization to
inspection of the resulting requirements. The framework of \citet{haley2008} derives
security requirements from the assets to be protected and the harm that threats could
cause to them, and checks with explicit satisfaction arguments that the requirements hold
under stated trust assumptions about the environment. Making such assumptions explicit, as
in \cref{eq:satisfaction}, is particularly important for security, because a violated
assumption can leave a requirement unsatisfied even when the software behaves as
specified \citep{haley2008}.
When a conflict between security and another quality cannot be avoided, the decision should
be based on the stakeholders' agreed priorities and the assessed risk, and it should be
recorded with its rationale.

\subsection{Prioritization}

Not all requirements can be satisfied within the available time and budget, so they must be
ranked. Prioritization weighs the value of each requirement to the stakeholders against its
cost and risk \citep{wiegers2013}; dependencies between requirements further constrain the
order in which they can be delivered. A value-based view of
RE \citep{aurum2007} makes this explicit by treating requirements decisions as choices that
should maximize the value the system delivers to its stakeholders. Iterative and agile
lifecycles depend on prioritization, because it determines what each iteration delivers. In
sequential lifecycles it determines what is cut when the schedule slips and which
requirements should receive the most careful validation.

% ===========================================================================
\section{Specification, Validation and Change}
\label{sec:spec}

\subsection{Writing requirements}

A well-written requirement has a single interpretation for all its readers. Requirements in
natural language are commonly written as ``shall'' statements, one requirement per
statement, in the vocabulary of the domain and the glossary
\citep{wiegers2013,iso29148}. Each requirement should be necessary (it adds value for some
stakeholder), feasible within the known constraints of the system and its environment,
verifiable by test, inspection, demonstration or analysis, and uniquely identified so that
it can be traced. Natural language can be supplemented by diagrams and models (use cases,
state machines, data models), and, for critical properties, by formal notations
\citep{vanlamsweerde2009}. Formality is not necessary in every case, but every statement
must still have a precise meaning.

\subsection{Validation}
\label{sec:validation}

Validation checks that the specified requirements describe what the stakeholders need and
that they can be implemented as stated. Common techniques are reviews and inspections of the
requirements document with stakeholders and developers, prototypes that let stakeholders
try the specified behavior, and the definition of acceptance criteria or test cases for
each requirement, which quickly exposes requirements that are ambiguous or unverifiable
\citep{nuseibeh2000,wiegers2013}. Validation is not a single step at the end of RE. It is
repeated whenever requirements are added or changed, so that errors are caught while they
are still cheap to fix (\cref{sec:cost}).

\subsection{Managing change}

Requirements change throughout a project as stakeholders learn more, the business
environment changes and technical constraints are discovered. Freezing the requirements does
not stop stakeholders' needs from changing; it only defers the point at which the change
must be dealt with. Change
should instead be managed: proposed changes are recorded, their impact on scope, schedule,
architecture and other requirements is analyzed, a responsible person or board decides, and
the plan and the affected stakeholders are updated. Traceability links from each requirement
to its source and to the design elements and tests that realize it make impact analysis
feasible. Controlled change in this way prevents uncontrolled scope growth (scope creep)
while allowing the requirements to improve over the project. \Cref{tab:challenges}
summarizes the main challenges discussed in this paper and the corresponding mitigations.

\begin{table}[t]
  \caption{Summary of common requirements challenges and mitigations.}
  \label{tab:challenges}
  \small
  \begin{tabularx}{\textwidth}{@{}P{3.6cm}L@{}}
    \toprule
    Challenge & Mitigation \\
    \midrule
    Limited access to stakeholders &
      Agree on communication channels and schedules early; explain how input will be used;
      use asynchronous techniques such as surveys and document analysis to supplement
      meetings. \\
    Tacit and unstated knowledge &
      Combine interviews with observation and prototypes; confirm information from
      several sources; record assumptions explicitly. \\
    Vague or unverifiable requirements &
      Ask clarifying questions; quantify quality requirements; define acceptance criteria
      for every requirement (\cref{tab:vague}). \\
    Conflicting stakeholder priorities &
      Make conflicts explicit; agree on shared project goals; negotiate alternatives;
      prioritize by value, cost and risk. \\
    Security versus performance and usability &
      State security requirements as protection goals for specific assets and threats;
      decide trade-offs on assessed risk and record the rationale. \\
    Scope creep &
      Baseline the requirements; apply a change-control process with impact analysis;
      update the plan and inform stakeholders. \\
    Too much or too little documentation &
      Match the documentation to the lifecycle, regulatory needs and audience; keep it
      current; document decisions and rationale even in agile projects. \\
    \bottomrule
  \end{tabularx}
\end{table}

% ===========================================================================
\section{Conclusion}
\label{sec:conclusion}

Requirements defects are a major source of project risk. Survey evidence, despite its
methodological weaknesses, consistently places incomplete requirements and lack of user
involvement among the leading causes of project trouble, and requirements errors that are
carried into design and implementation can be expensive to correct. The practices described in this paper address these risks at their source:
understanding the domain before specifying the system, involving a representative range of
stakeholders, distinguishing requirements from domain properties, assumptions and
definitions, turning vague expectations (including security expectations) into verifiable
requirements, validating continuously, and managing change rather than resisting it.

No lifecycle model removes the need for these activities. Sequential models concentrate RE
at the start of a project and depend on getting it right early, whereas iterative and agile
models spread it across the project and depend on continuous stakeholder involvement and
careful prioritization. The choice of model changes how much is written down and when, but
not the underlying goal of building a system that does what its stakeholders need.

This paper is a narrative overview. It does not systematically review the literature, and
the quantitative evidence it cites is dated and contested; readers who need firmer evidence
should consult recent systematic studies of RE practice.

\renewcommand{\bibfont}{\small}
\setlength{\bibsep}{3pt plus 1pt}
\bibliographystyle{plainnat}
\bibliography{references}

\end{document}

%% file: figures/process.tex
% Requirements engineering activities (TikZ)
\begin{tikzpicture}[
    font=\small,
    act/.style={draw=accent, line width=0.7pt, fill=accentlight, rounded corners=2pt,
                text width=3.15cm, minimum height=1.45cm, align=center, text=ink},
    io/.style={draw=muted, line width=0.5pt, rounded corners=2pt, fill=white,
               text width=4.0cm, minimum height=0.8cm, align=center, text=ink,
               font=\footnotesize},
    flow/.style={-{Stealth[length=2.2mm]}, line width=0.7pt, draw=ink},
    back/.style={-{Stealth[length=2.2mm]}, line width=0.6pt, draw=muted, dashed},
    note/.style={font=\footnotesize\itshape, text=muted, fill=white, inner sep=2pt}
  ]
  % core activities
  \node[act] (eli) at (0,0)     {Domain\\understanding\\and elicitation};
  \node[act] (ana) at (3.8,0)  {Analysis and negotiation};
  \node[act] (spe) at (7.6,0)   {Specification};
  \node[act] (val) at (11.4,0) {Validation};
  \draw[flow] (eli) -- (ana);
  \draw[flow] (ana) -- (spe);
  \draw[flow] (spe) -- (val);

  % input and output
  \node[io] (src) at (0,1.75)     {Stakeholders, documents,\\existing system};
  \node[io] (srs) at (11.4,1.75) {Agreed requirements\\(SRS, use cases, backlog)};
  \draw[flow] (src) -- (eli);
  \draw[flow] (val) -- (srs);

  % feedback loop below the row
  \draw[back] (val.south) -- ++(0,-0.6) -| node[note, pos=0.25]
     {feedback: open questions, misunderstandings, change requests} (eli.south);

  % requirements management band
  \node[draw=muted, line width=0.5pt, fill=rule, rounded corners=2pt,
        minimum width=14.6cm, minimum height=0.62cm, font=\footnotesize, text=ink]
        (mgmt) at (5.7,-1.85)
        {Requirements management throughout the project: change control, versioning, traceability};
\end{tikzpicture}

%% file: figures/worldmachine.tex
% Environment / software / shared phenomena: train door example (TikZ)
\begin{tikzpicture}[
    font=\small,
    stmt/.style={draw=ink, line width=0.5pt, fill=white, rounded corners=2pt, align=left,
                 text width=#1, inner sep=4pt, font=\footnotesize, text=ink, anchor=north west},
    tag/.style={font=\scriptsize\bfseries, text=muted, anchor=south west, inner sep=1pt},
    lnk/.style={line width=0.55pt, draw=muted, -{Stealth[length=1.8mm]}},
    ph/.style={font=\footnotesize\ttfamily, text=ink, fill=white, draw=accent,
               line width=0.5pt, rounded corners=1.5pt, inner sep=3pt}
  ]
  % regions
  \filldraw[fill=rule, draw=muted, line width=0.6pt, rounded corners=4pt] (0,0) rectangle (8.2,6.0);
  \filldraw[fill=accentlight, draw=accent, line width=0.6pt, rounded corners=4pt] (10.6,0) rectangle (14.6,6.0);
  \draw[accent, line width=0.6pt, dashed] (8.35,0) -- (8.35,6.0);
  \draw[accent, line width=0.6pt, dashed] (10.45,0) -- (10.45,6.0);

  \node[font=\small\bfseries, text=ink, anchor=north west] at (0.2,5.85) {Environment};
  \node[font=\small\bfseries, text=ink, anchor=north east] at (14.4,5.85) {Software-to-be};
  \node[font=\small\bfseries, text=ink, align=center, anchor=north] at (9.4,5.85) {Shared\\phenomena};

  % statements
  \node[stmt=7.6cm] (sys) at (0.25,4.85)
     {Doors shall remain closed while the train is moving.};
  \node[tag] at ([yshift=1pt]sys.north west) {SysReq (prescriptive)};

  \node[stmt=3.3cm] (dom) at (0.25,3.35)
     {A train is moving if and only if its physical speed is non-zero.};
  \node[tag] at ([yshift=1pt]dom.north west) {Dom (descriptive)};

  \node[stmt=3.55cm] (asm1) at (4.25,3.35)
     {Sensor: \texttt{measuredSpeed}${}\neq 0$ iff physical speed${}\neq 0$.};
  \node[tag] at ([yshift=1pt]asm1.north west) {Asm (on environment)};

  \node[stmt=3.55cm] (asm2) at (4.25,1.3)
     {Actuator: \mbox{\texttt{doorsState}${}={}$\texttt{closed}} iff the doors are closed.};
  \node[tag] at ([yshift=1pt]asm2.north west) {Asm (on environment)};

  \node[stmt=3.3cm] (sof) at (10.95,3.35)
     {\mbox{\texttt{doorsState}${}={}$\texttt{closed}} whenever \texttt{measuredSpeed}${}\neq 0$.};
  \node[tag] at ([yshift=1pt]sof.north west) {SofReq (prescriptive)};

  % shared phenomena
  \node[ph] (ms) at (9.4,2.75) {measuredSpeed};
  \node[font=\scriptsize, text=muted, below=0pt of ms] {input};
  \node[ph] (ds) at (9.4,0.75) {doorsState};
  \node[font=\scriptsize, text=muted, below=0pt of ds] {output};

  \draw[lnk] (asm1.east |- ms) -- (ms.west);
  \draw[lnk] (ds.west) -- (asm2.east |- ds);
  \draw[lnk] (ms.east) -- (sof.west |- ms);
  \draw[lnk] (sof.south) |- (ds.east);
\end{tikzpicture}

%% file: figures/chaos.tex
% CHAOS 1994 results (pgfplots). Data: The Standish Group (1994).
\pgfplotsset{
  chaosbase/.style={
    xbar, xmin=0, point meta=x, axis lines*=left, axis line style={draw=muted, line width=0.5pt},
    y axis line style={draw=none}, ytick style={draw=none},
    xtick style={draw=muted, line width=0.5pt},
    tick label style={font=\footnotesize, text=ink},
    x tick label style={text=muted},
    xmajorgrids, major grid style={draw=rule, line width=0.4pt},
    y dir=reverse, enlarge y limits={abs=0.55},
    nodes near coords, nodes near coords style={font=\scriptsize, text=ink, anchor=west},
    label style={font=\footnotesize, text=muted},
    title style={font=\small, text=ink, anchor=west, at={(-0.0,1.0)}, xshift=0pt, align=left},
    clip=false,
  }
}
\begin{tikzpicture}
  % (a) project resolution
  \begin{axis}[chaosbase, name=a,
      width=10.2cm, height=3.3cm, scale only axis=false,
      xmax=60, xtick={0,10,...,60}, bar width=7pt,
      ytick={0,1,2},
      yticklabels={Successful,Challenged,{Impaired (canceled)}},
      xlabel={Share of projects (\%)},
      nodes near coords={\pgfmathprintnumber[fixed,zerofill,precision=1]{\pgfplotspointmeta}\%},
      title={\textbf{(a)} Project resolution},
      title style={at={(-0.31,1.04)}}]
    \addplot[fill=muted, draw=none] coordinates
      {(16.2,0) (52.7,1) (31.1,2)};
  \end{axis}
\end{tikzpicture}

\vspace{4pt}
\begin{tikzpicture}
  % (b) challenged factors
  \begin{axis}[chaosbase,
      width=3.3cm, height=4.0cm, scale only axis,
      xmax=15, xtick={0,5,10,15}, bar width=6.5pt, bar shift=0pt,
      ytick={0,...,4},
      yticklabels={Lack of user input,{Incomplete requirements\\and specifications},{Changing requirements\\and specifications},Lack of executive support,Technology incompetence},
      y tick label style={align=right, font=\footnotesize},
      xlabel={Share of responses (\%)},
      nodes near coords={\pgfmathprintnumber[fixed,zerofill,precision=1]{\pgfplotspointmeta}},
      title={\textbf{(b)} Challenged projects},
      title style={at={(-1.12,1.06)}}]
    \addplot[fill=accent, draw=none] coordinates
      {(12.8,0) (12.3,1) (11.8,2)};
    \addplot[fill=neutral, draw=none] coordinates
      {(7.5,3) (7.0,4)};
  \end{axis}
\end{tikzpicture}\hspace{0.4cm}%
\begin{tikzpicture}
  % (c) impaired factors
  \begin{axis}[chaosbase,
      width=3.3cm, height=4.0cm, scale only axis,
      xmax=15, xtick={0,5,10,15}, bar width=6.5pt, bar shift=0pt,
      ytick={0,...,4},
      yticklabels={Incomplete requirements,Lack of user involvement,Lack of resources,Unrealistic expectations,Lack of executive support},
      y tick label style={align=right, font=\footnotesize},
      xlabel={Share of responses (\%)},
      nodes near coords={\pgfmathprintnumber[fixed,zerofill,precision=1]{\pgfplotspointmeta}},
      title={\textbf{(c)} Impaired projects},
      title style={at={(-1.05,1.06)}}]
    \addplot[fill=accent, draw=none] coordinates
      {(13.1,0) (12.4,1)};
    \addplot[fill=neutral, draw=none] coordinates
      {(10.6,2) (9.9,3) (9.3,4)};
  \end{axis}
\end{tikzpicture}

\vspace{3pt}
{\footnotesize
\tikz{\fill[accent] (0,0) rectangle (0.32,0.2);}\ Requirements or user involvement
\qquad
\tikz{\fill[neutral] (0,0) rectangle (0.32,0.2);}\ Other factor}

%% file: figures/lifecycles.tex
% Schematic distribution of RE effort over time in four lifecycle models (pgfplots).
% Curves are illustrative, not measured data.
\pgfmathdeclarefunction{g}{3}{\pgfmathparse{#3*exp(-((x-#1)/#2)^2)}}
\begin{tikzpicture}
  \begin{groupplot}[
      group style={group size=1 by 4, vertical sep=0.35cm, xlabels at=edge bottom},
      width=11.2cm, height=1.3cm, scale only axis,
      xmin=0, xmax=1, ymin=0, ymax=1.25,
      axis lines*=left, axis line style={draw=muted, line width=0.5pt},
      ytick=\empty, xtick=\empty,
      re/.style={line width=0.8pt, draw=accent, fill=accentlight, mark=none,
                 samples=241, domain=0:1, smooth},
      ylabel style={font=\footnotesize, text=ink, rotate=-90, anchor=east, align=right},
      clip=false,
    ]
    % Waterfall
    \nextgroupplot[ylabel={Waterfall}]
    \addplot[re, sharp plot] coordinates {(0,0) (0.005,1) (0.2,1) (0.205,0) (1,0)} \closedcycle;
    \node[font=\scriptsize, text=muted, anchor=west] at (axis cs:0.22,0.62)
      {requirements phase; later changes via formal change control};

    % Spiral
    \nextgroupplot[ylabel={Spiral}]
    \addplot[re] {g(0.04,0.05,1.0)+g(0.29,0.05,0.75)+g(0.54,0.05,0.55)+g(0.79,0.05,0.38)} \closedcycle;
    \draw[muted, dashed, line width=0.4pt] (axis cs:0.25,0) -- (axis cs:0.25,1.22);
    \draw[muted, dashed, line width=0.4pt] (axis cs:0.5,0) -- (axis cs:0.5,1.22);
    \draw[muted, dashed, line width=0.4pt] (axis cs:0.75,0) -- (axis cs:0.75,1.22);
    \node[font=\scriptsize, text=muted] at (axis cs:0.15,1.1) {cycle 1};
    \node[font=\scriptsize, text=muted] at (axis cs:0.4,1.1) {cycle 2};
    \node[font=\scriptsize, text=muted] at (axis cs:0.65,1.1) {cycle 3};
    \node[font=\scriptsize, text=muted] at (axis cs:0.9,1.1) {cycle 4};

    % RUP  (phase split 10/30/50/10)
    \nextgroupplot[ylabel={RUP}]
    \addplot[re] {0.05 + g(0.13,0.12,0.88) + g(0.03,0.05,0.35) + g(0.55,0.25,0.18) + g(0.47,0.02,0.06) + g(0.6,0.02,0.06) + g(0.73,0.02,0.05)} \closedcycle;
    \draw[muted, dashed, line width=0.4pt] (axis cs:0.1,0) -- (axis cs:0.1,1.22);
    \draw[muted, dashed, line width=0.4pt] (axis cs:0.4,0) -- (axis cs:0.4,1.22);
    \draw[muted, dashed, line width=0.4pt] (axis cs:0.9,0) -- (axis cs:0.9,1.22);
    \node[font=\scriptsize, text=muted] at (axis cs:0.05,1.16) {I};
    \node[font=\scriptsize, text=muted] at (axis cs:0.25,1.16) {E};
    \node[font=\scriptsize, text=muted] at (axis cs:0.65,1.16) {C};
    \node[font=\scriptsize, text=muted] at (axis cs:0.95,1.16) {T};

    % Agile
    \nextgroupplot[ylabel={Agile}, xlabel={Project time (normalized)},
                   xlabel style={font=\footnotesize, text=muted},
                   xtick={0,1}, xticklabels={start,end}, x tick label style={font=\scriptsize, text=muted}]
    \addplot[re] {0.08 + g(0.03,0.035,0.55)
              + g(0.105,0.018,0.32)+g(0.205,0.018,0.32)+g(0.305,0.018,0.32)+g(0.405,0.018,0.32)
              + g(0.505,0.018,0.32)+g(0.605,0.018,0.32)+g(0.705,0.018,0.32)+g(0.805,0.018,0.32)
              + g(0.905,0.018,0.32)} \closedcycle;
    \node[font=\scriptsize, text=muted, anchor=west] at (axis cs:0.55,0.95)
      {backlog refinement in every iteration};
  \end{groupplot}
\end{tikzpicture}